\documentclass[a4paper,11pt]{article}
\pdfoutput=1
\usepackage{jheppub}
\usepackage{amsmath,amssymb,bm,graphicx,bbold,epsf,colordvi}
\usepackage{lipsum}
\usepackage{makecell}
\usepackage{caption}
\usepackage{subcaption}
\usepackage{sidecap}
\usepackage{soul}
\usepackage{braket}
\usepackage{bm}
\usepackage{appendix}
\newcommand{\jet}{{J}}
\allowdisplaybreaks 
\usepackage{color}
\usepackage[dvipsnames]{xcolor}

\usepackage{mathrsfs}
\usepackage{slashed}
\allowdisplaybreaks
\usepackage[abs]{overpic}
\usepackage[export]{adjustbox}
\usepackage{bm}
\usepackage{scalerel}
\usepackage{accents}

\newcommand{\bef}{\begin{figure}[hbt]\centering}
\newcommand{\eef}{\end{figure}}
\usepackage{mathtools}
\usepackage{booktabs}
\usepackage{graphicx}
\graphicspath{ {./figure/} }
\usepackage{comment}
\usepackage{lipsum}

\newcommand{\nnu}{\nonumber\\}

\newcommand{\beq}{\begin{equation}}
\newcommand{\eeq}{\end{equation}}
\def\bea#1\eea{\begin{align}#1\end{align}}

\def \be  {\begin{equation}}
\def \ee  {\end{equation}}
\def \ba  {\begin{eqnarray}}
\def \ea  {\end{eqnarray}}

\allowdisplaybreaks

\newenvironment{Presented}{\begin{quotation} \begin{center} 
             PRESENTED AT\end{center}\bigskip 
      \begin{center}\begin{large}}{\end{large}\end{center} \end{quotation}}

\makeatletter
\def\@fpheader{
\begin{Presented}
DIS2023: XXX International Workshop on Deep-Inelastic Scattering and
Related Subjects, \\
Michigan State University, USA, 27-31 March 2023 \\
     \includegraphics[width=9cm]{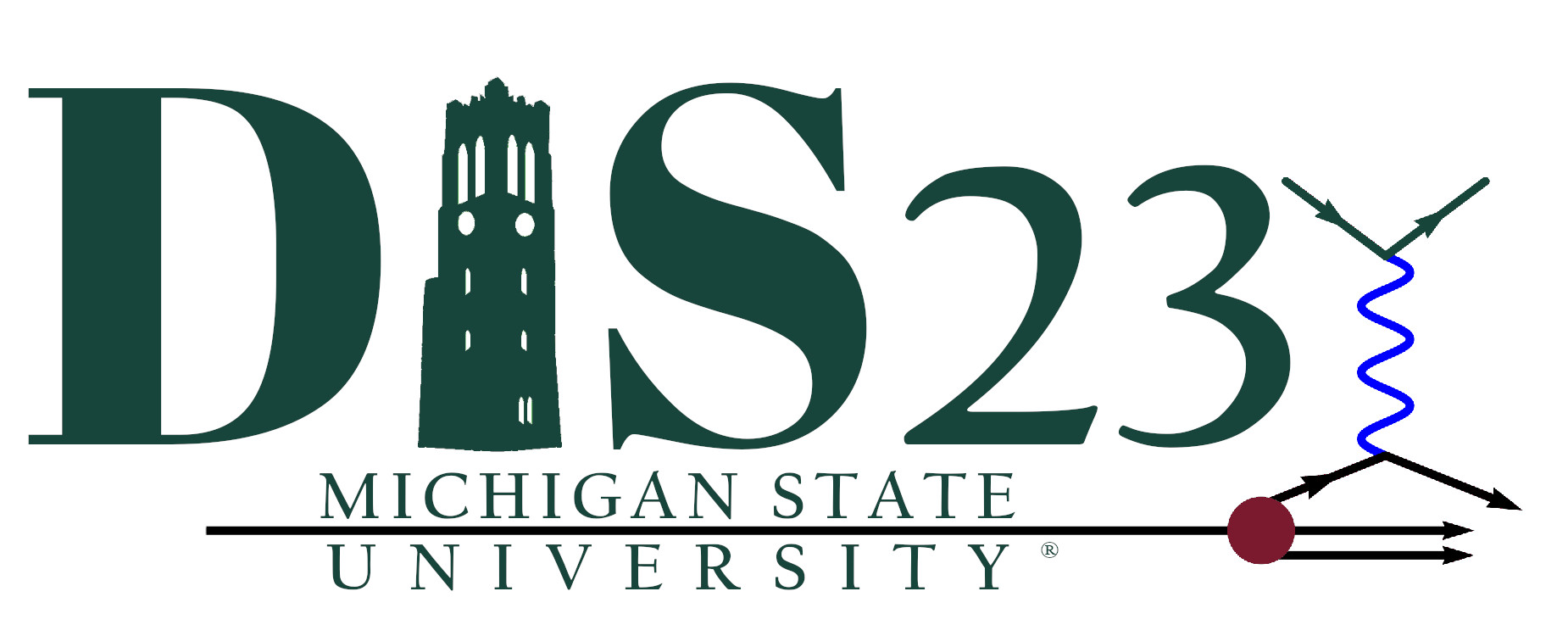}
\end{Presented}
}
\makeatother

\usepackage[Q=yes,pverb-linebreak=no]{examplep}

\title{Collins-type Energy-Energy Correlators and Nucleon Structure}
\author[a,b,c]{Zhong-Bo Kang}
\author[d]{, Kyle Lee}
\author[e,f]{, Ding Yu Shao}
\author[a,b]{and Fanyi Zhao}

\affiliation[a]{Department of Physics and Astronomy, University of California, Los Angeles, CA 90095, USA}
\affiliation[b]{Mani L. Bhaumik Institute for Theoretical Physics, University of California, Los Angeles, CA 90095, USA}
\affiliation[c]{Center for Frontiers in Nuclear Science, Stony Brook University, Stony Brook, NY 11794, USA}
\affiliation[d]{Center for Theoretical Physics, Massachusetts Institute of Technology, Cambridge, MA 02139}
\affiliation[e]{Department of Physics and Center for Field Theory and Particle Physics, Fudan University, Shanghai, China}
\affiliation[f]{Key Laboratory of Nuclear Physics and Ion-beam Application (MOE), Fudan University, Shanghai, China}

\emailAdd{zkang@ucla.edu, kylel@mit.edu, dingyu.shao@cern.ch, fanyizhao@physics.ucla.edu}

\abstract
{We generalize the conventional Energy-Energy Correlator (EEC) to include the azimuthal angle dependence, so to define azimuthal angle dependent EEC observables. We study this new EEC observable in $e^+e^-$ and semi-inclusive deep inelastic scattering (SIDIS). In the back-to-back region, we find that the azimuthal angle dependent EEC is sensitive to both the unpolarized EEC jet function and a Collins-type EEC jet function. While the unpolarized EEC jet function is related to the unpolarized transverse momentum dependent (TMD) fragmentation function, the Collins-type EEC jet function is connected with the Collins fragmentation function. We further demonstrate how the new observables allow us to access to the 3D structure of nucleons, especially the spin-dependent ones.}

\begin{document}
\maketitle

\section{Introduction}\label{sec:intro}
The energy-energy correlators (EEC)~\cite{Basham:1978bw,Basham:1978zq} is one of the earliest infrared and collinear (IRC) safe event shape observables~\cite{Kinoshita:1962ur,Lee:1964is}. This event shape observable has been studied with unpolarized scattering beams in electron-positron collisions and Semi-Inclusive Deep Inelastic Scattering (SIDIS). Various experimental measurements have been provided for $e^+e^-$ collisions~\cite{CELLO:1982rca,JADE:1984taa,Fernandez:1984db,PLUTO:1985yzc,TASSO:1987mcs,Wood:1987uf}, which have revealed insights into the structure and properties of hadrons and other subatomic particles. 

\begin{figure}[h]
\centering
\includegraphics[height=0.245\textwidth]{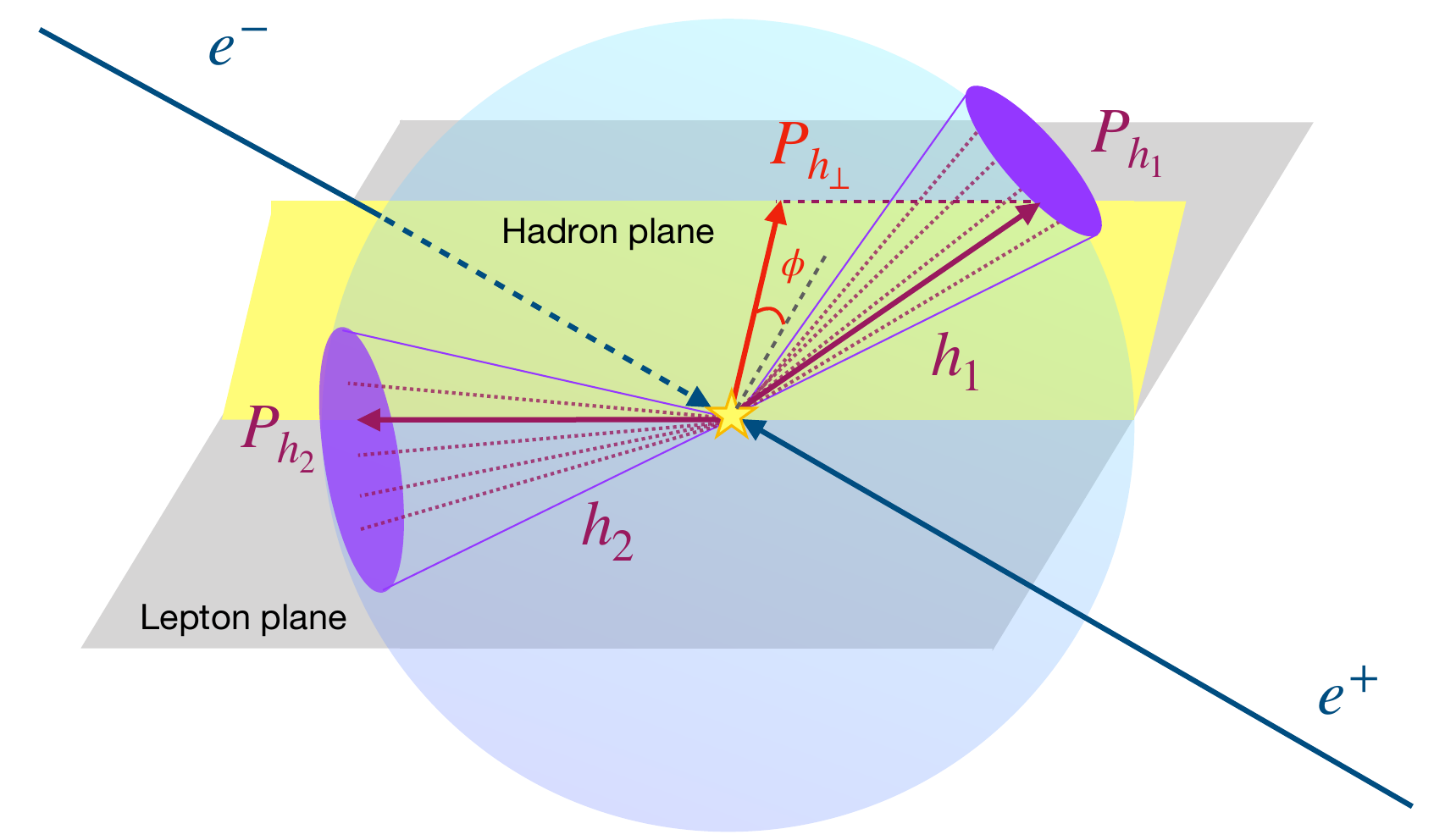}\quad
\includegraphics[height=0.245\textwidth]{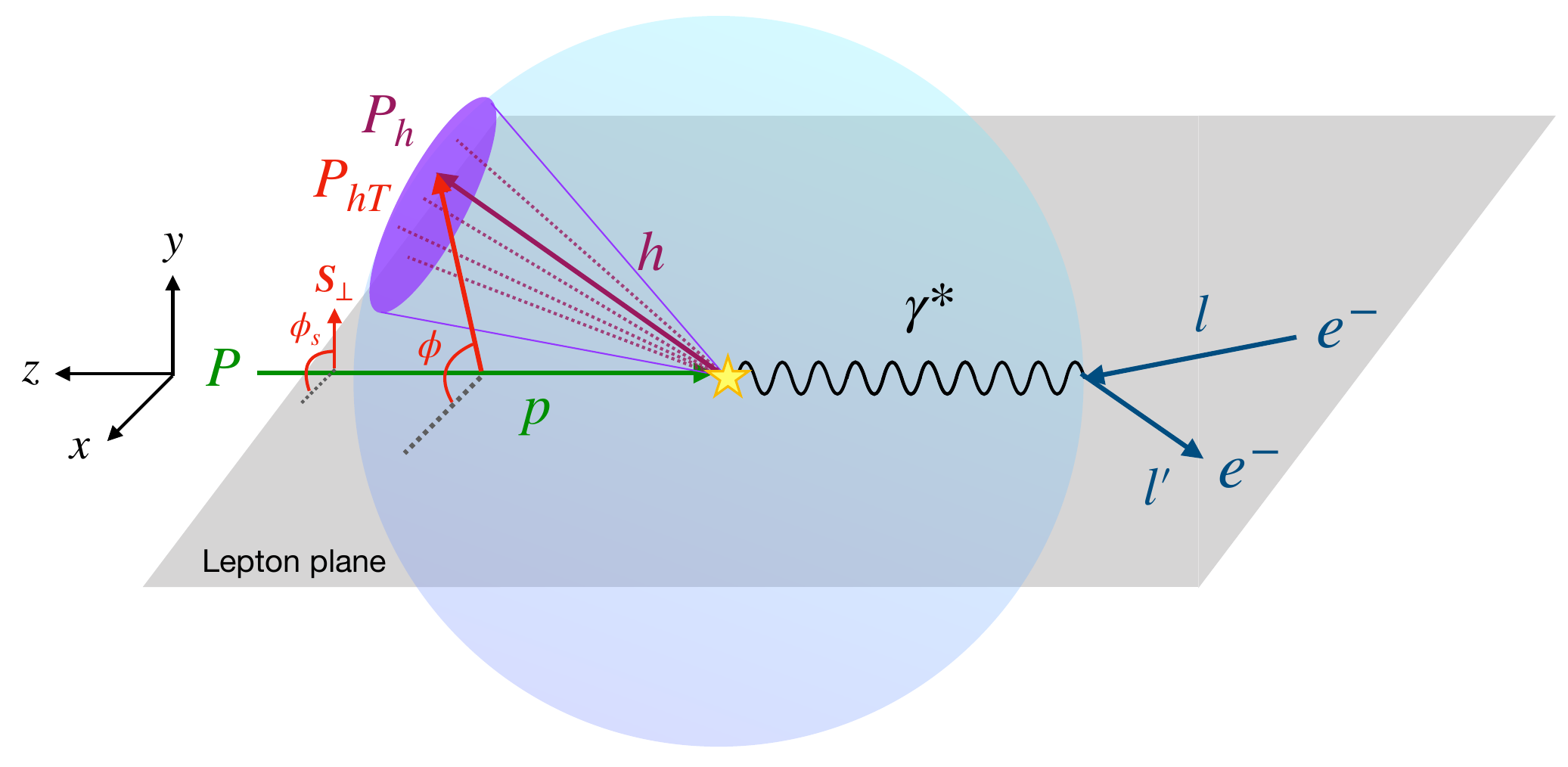}
\caption{Illustration of EEC for $e^+e^-$ annihilation in GJ frame (left) and for Deep Inelastic Scattering in Breit frame (right).}\label{fig:eec_ee}
\end{figure}

The conventional EEC observable measures the correlation between the energies of two particles produced in the collision, while they are separated by the angle $\theta_{ij}$~\cite{Moult:2018jzp}. In this work, we generalize this observable to include the azimuthal angle $\phi_{ij}$ between the two particles and study how this azimuthal angle dependent EEC enables us to go beyond what is possible with the conventional EEC for probing nucleon structure, in particular the spin-dependent nucleon structure.  We start with the definition of this observable in $e^+e^-$ collisions, 
\begin{align}
\label{eq:EECazdef}
\mathrm{{{EEC}}}_{e^+e^-}(\tau,\phi)=&\frac{1}{2} \sum_{i,j}\int d\theta_{ij} dz_i dz_j z_i z_j \frac{1}{\sigma} \frac{d\sigma}{dPS} \delta\left(\tau - \left(\frac{1+\cos \theta_{ij}}{2}\right)\right) \delta(\phi - \phi_{ij})\,,
\end{align}
where $z_i$ and $z_j$ are the energy fractions of the final-state hadrons $i,j$, separated by $\theta_{ij}$, angle $\phi_{ij}$ is the azimuthal angle difference between the two hadrons in the Gottfried-Jackson (GJ) frame\ \cite{Gottfried:1964nx} as illustrated in the left panel of Fig.\ \ref{fig:eec_ee}, and $dPS=d\theta_{ij}d\phi_{ij} dz_i dz_j$. In the next section where we study the EEC in the back-to-back region, we will demonstrate that this azimuthal angle dependent polarized EEC gives sensitivity to both the traditional EEC jet function as well as a new EEC jet function, named ``Collins-type'' EEC jet function. The traditional EEC jet function has a close relation to the unpolarized fragmentation functions $D_{1}(z, {k}_\perp^2)$~\cite{Moult:2018jzp}, while the Collins-type EEC jet function is  related to the well-known Collins fragmentation functions $H_{1}^\perp(z,{k}_\perp^2)$~\cite{toappear}. In the section below, we will also discuss the EEC for the SIDIS. 

\section{EEC in the back-to-back region for $e^+e^-$ annihilation}
In the back-to-back region, one has $\theta_{ij}\to \pi$ and thus $\tau\to 0$ in Eq.~\eqref{eq:EECazdef}. It can be shown $\displaystyle \tau = {\bm{P}_{h_\perp}^2}/({z_1^2Q^2})$ in this region. 
Here $\bm{P}_{h_\perp}$ is the transverse momentum of the hadron $h_1$ in the so-called Gottfried-Jackson (GJ) frame\ \cite{Gottfried:1964nx}, namely with respect to hadron $h_2$ in the pair production, see Fig.\ \ref{fig:eec_ee}.
By introducing $\bm{q}_T = -\bm{P}_{h_\perp}/z_1$, one obtains the observable $\tau$ represented by $\displaystyle \tau = {\bm{q}_{T}^2}/{Q^2}$.

\begin{SCfigure}
\includegraphics[height=0.32\textwidth]{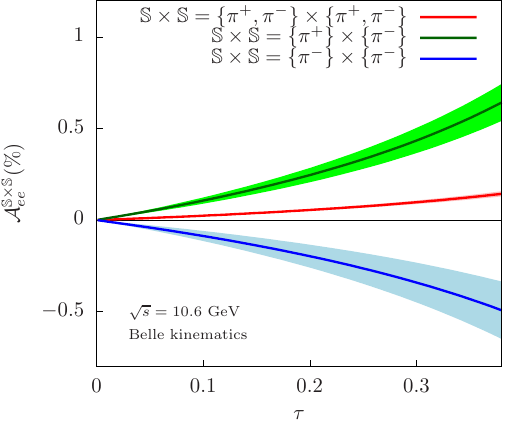}
\caption{$A_{ee}^{\mathbb{S}\times\mathbb{S}}(\tau)$ for $\mathbb{S}\times\mathbb{S}=\{\pi^+,\ \pi^-\}\times\{\pi^+,\ \pi^-\}$, $\{\pi^+\}\times\{\pi^-\}$ and $\{\pi^-\}\times\{\pi^-\}$ at $\sqrt{s}=10.6$ GeV with Collins functions fitted in~\cite{Kang:2015msa}.} \label{fig:asye+e-}    
\end{SCfigure}

Using these variables to rewrite Eq.~\eqref{eq:EECazdef}, we find that the azimuthal angle dependent EEC, $\mathrm{{{EEC}}}_{e^+e^-}(\tau,\phi)$, is related to $\bm{q}_T$-differential cross section as 
\begin{align}
\label{eq:EECazdef2}
\mathrm{{{EEC}}}_{e^+e^-}(\tau,\phi) &=\frac{1}{2} \sum_{i,j}\int d^2\bm{q}_T dz_i dz_j z_i z_j \frac{1}{\sigma} \frac{d\sigma}{d^2\bm{q}_T dz_i dz_j} \delta\left(\tau - \frac{\bm{q}_T^2}{Q^2}\right) \delta(\phi - \phi_{q_T})\,.
\end{align}
Starting from the azimuthal angle dependent factorization in the GJ frame (see e.g. in \cite{Kang:2015msa}), especially after Fourier transforming from $\bm q_T$-space to $\bm b$-space, one would end up with the usual unpolarized EEC jet function $\jet_q({\bm b},\mu,\zeta)$~\cite{Moult:2018jzp} and the Collins-type EEC jet functions $\jet_q^\perp({\bm b},\mu,\zeta)$
\begin{align}
\label{eq:unpjet}
\hspace{-0.2cm}\jet_q({\bm b},\mu,\zeta)\equiv&\sum_{h}\int_0^1dz\,z\,\tilde{D}_{1,h/q}(z,\mu_{b_*},\zeta_i) e^{-S_{\rm pert}(\mu,\mu_{b_*}) - S_{\rm NP}^{D_1}(b,Q_0,\zeta)}\left(\sqrt{\frac{\zeta}{\zeta_i}}\right)^{\kappa\left(b, \mu_{b_*}\right)}\,,\\
\hspace{-0.2cm}\jet_q^\perp({\bm b},\mu,\zeta)\equiv&\sum_{h}\int_0^1dz\,\tilde{H}_{1,h/{q}}^{\perp}(z,\mu_{b_*},\zeta_i)e^{-S_{\rm pert}(\mu,\mu_{b_*}) - S_{\rm NP}^{H^{\perp}_{1}}(b,Q_0,\zeta)}\left(\sqrt{\frac{\zeta}{\zeta_i}}\right)^{\kappa\left(b, \mu_{b_*}\right)}\label{eq:Collinsjet}\,,
\end{align}
where $\tilde{D}_{1,h/q}$ and $\tilde{H}_{1,h/{q}}^{\perp}$ are related to unpolarized TMD FFs and Collins functions respectively. At the end of the day, we obtain the $\phi$-dependent EEC,
\begin{align}
\label{eq:epemEEC}
\mathrm{{{EEC}}}_{e^+e^-}(\tau,\phi)=&\,\frac{2\pi N_c\alpha_{\rm em}^2}{3Q^2\sigma}\sum_q e_q^2\int d{\bm q}_T^2\delta(\tau  - \frac{{\bm q}_T^2}{Q^2})\left(Z_{uu}+\cos 2\phi\,Z_{\rm collins}\right)\,,
\end{align}
where we have 
\begin{align}
Z_{uu}=&\int\frac{bdb}{2\pi}J_0(bq_T)\jet_{q}({\bm b},\mu,\zeta)\jet_{\bar{q}}({\bm b},\mu,\zeta)\,\\
Z_{\rm collins}=&\,\int\frac{bdb}{2\pi}\frac{b^2}{8}J_2(bq_T)\jet_{q}^{\perp}({\bm b},\mu,\zeta)\jet_{\bar{q}}^{\perp}({\bm b},\mu,\zeta)
\end{align}
For a phenomenology example, we study the ratio $A_{ee}(\tau)={Z_{\rm collins}}\big/{Z_{uu}}$. Moreover, by substituting the summation over all the final hadrons $h$ in Eqs.~\eqref{eq:unpjet} and~\eqref{eq:Collinsjet} to a subset $\mathbb{S}$ for each direction in the pair production, one has EEC of subsets $\mathbb{S}\times\mathbb{S}$. As pions are the primary measurements available for the Collins-type asymmetry, we now study the ratio $A_{ee}^{\mathbb{S}\times\mathbb{S}}(\tau) $ with different combinations of pion pairs. In Fig.~\ref{fig:asye+e-}, we plot the Collins asymmetry for subsets $\mathbb{S}=\{\pi^+\},\ \{\pi^-\}$ or $\{\pi^\pm\}$ using TMD FFs $D_1(z,k_\perp^2)$ and Collins function $H_1^\perp(z,k_\perp^2)$ extracted in~\cite{Kang:2015msa}. When choosing a subset of either positively or negatively charged pions detected in EEC, one observes sizable asymmetries, which worth further measurement.
\section{EEC in the back-to-back region for SIDIS}
\begin{figure}[h]
\centering
\includegraphics[height=0.33\textwidth]{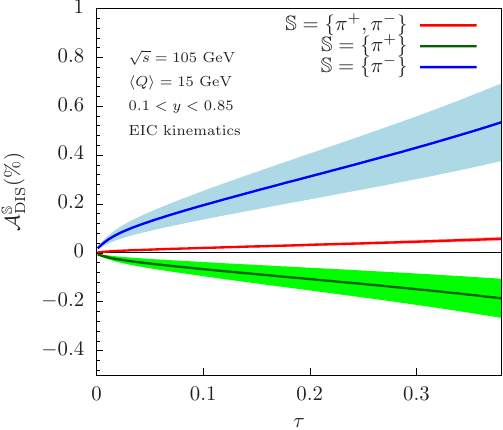}\hspace{1.5cm}
\includegraphics[height=0.33\textwidth]{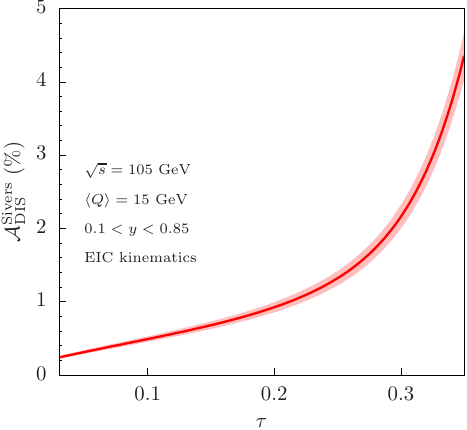}
\caption{$A_{\rm DIS}^{\mathbb{S}=\{\pi^\pm\},\ \{\pi^+,\ \pi^-\}}$ (left) and $A_{\rm DIS}^{\rm Sivers}$ for $\mathbb{S}=\{\pi^+,\ \pi^-\}$ (right) at EIC kinematics. }\label{fig:asyDIS}
\end{figure}

SIDIS is a fundamental tool in the study of the structure of nucleons and nuclei. For the process
\begin{align}
e^-(l)+p(P)\rightarrow e^-(l')+h(P_h)+X\,,
\end{align}
the SIDIS version of the EEC that will be generalized to {{EEC}}$_{\rm DIS}$ here was developed recently~\cite{Li:2021txc}, and was characterized as a function of the angular separation $\theta_{ap}$ between the incoming beam $p$ and the outgoing hadron $a$. We we further generalize it to the azimuthal angle dependent case as shown in the right panel of Fig.~\ref{fig:eec_ee},
\begin{align}
\label{eq:EECazimDISdef}
\mathrm{{{EEC}}_{DIS}}(\tau,\phi) &=\frac{1}{2} \sum_{a}\int d\theta_{ap}d\phi_{ap} dz_a z_a \frac{1}{\sigma} \frac{d\sigma}{dPS_{\rm DIS}} \delta\left(\tau - \left(\frac{1+\cos \theta_{ap}}{2}\right)\right)\delta(\phi - \phi_{ap})\,.
\end{align}
Here $dPS_{\rm DIS}=d\theta_{ap} d\phi_{ap} dz_a$. Similarly, in the back-to-back limit ($\tau\to 0$), we find $\displaystyle \tau = {\bm{P}_{aT}^2}/({z_a^2 Q^2})$, where $Q = \sqrt{-q^2}$ is the invariant mass of the exchanged virtual photon, $\bm{P}_{aT}$ is the transverse momentum of the outgoing hadron measured with respect to the photon-beam axis. By introducing $\bm{q}_T \equiv -\bm{P}_{aT}/z_a$, $\mathrm{{{EEC}}_{DIS}}(\tau,\phi)$ can be related to the $\bm{q}_T$-differential cross-section in the back-to-back limit as
\begin{align}
\label{eq:EECazimDISdef2}
\mathrm{{{EEC}}}_{\rm DIS}(\tau,\phi) =  \frac{1}{\sigma}\frac{d\Sigma_{\rm DIS}}{d\tau d\phi} &=\frac{1}{2} \sum_{a}\int d^2\bm{q}_T dz_a z_a \frac{1}{\sigma} \frac{d\sigma}{d^2\bm{q}_T dz_a} \delta\left(\tau - \frac{\bm{q}_T^2}{Q^2}\right) \delta(\phi - \phi_{q_T})\,.
\end{align}
Starting from the relevant azimuthal angle dependent $\bm{q}_T$ factorization for the DIS process given in~\cite{Gourdin:1972kro,Kotzinian:1994dv,Diehl:2005pc,Bacchetta:2006tn,Anselmino:2008sga,Collins:2011zzd,Kang:2015msa}, using the EEC jet functions defined above in Eqs.\ \eqref{eq:unpjet} and\ \eqref{eq:Collinsjet}, the {{EEC}} for SIDIS can be written as
\begin{align}
\frac{d\Sigma_{\rm DIS}}{dxdyd\tau d\phi}=&\frac{2\pi\alpha_{\rm em}^2}{Q^2}\frac{1+(1-y)^2}{y}\int d^2{\bm q}_T\delta\left(\tau -\frac{{\bm q}_T^2}{Q^2}\right)\delta(\phi - \phi_{q_T})\int\frac{db\ b}{2\pi}\bigg\{\mathcal{F}_{UU}\nnu
&+\cos(2\phi_{q_T})\frac{2(1-y)}{1+(1-y)^2}\mathcal{F}_{UU}^{\cos(2\phi_{q_T})}+|{\bm S}_\perp|\sin(\phi_{q_T}-\phi_s)\mathcal{F}_{UT}^{\sin(\phi_{q_T}-\phi_s)}+\cdots\bigg\}\,,\label{eq:EECDIS3}
\end{align}
where $\mathcal{F}_{UU}\sim f_1\otimes J_q,\ \mathcal{F}_{UU}^{\cos\left(2\phi_{q_T}\right)}\sim h_1\otimes J_q^\perp,\ \mathcal{F}_{UT}^{\sin\left(\phi_{q_T}-\phi_s\right)}\sim f_{1T}^{\perp}\otimes J_q$, etc. For the polarizations, $|{\bm S}_\perp|$ is the transverse spin of the incoming proton, $\lambda_e$ is the helicity of the incoming electron, and the indices $A$ and $B$ of $\mathcal{F}_{AB}$ represent the polarization of the incoming electron and proton, respectively. When studying a subset $\mathbb{S}$ of the produced hadrons, one has the structure function given in the form of $\mathcal{F}_{AB}^{\mathbb{S}}$, with an extra superscript $\mathbb{S}$. Additionally, the angles $\phi_s$ and $\phi_{q_T}$ are the azimuthal angles of the ${\bm S}_\perp$ and $\bm{q}_T$. For phenomenology, we plot the Collins asymmetry $\mathcal{A}_{\rm DIS}^{\mathbb{S}}$ (left panel) and the Sivers asymmetry $\mathcal{A}_{\rm DIS}^{\rm Sivers}$ (right panel) in Fig.~\ref{fig:asyDIS}, where
\begin{align}
    &\mathcal{A}_{\rm DIS}^{\mathbb{S}}=\frac{\mathcal{F}_{UU}^{\mathbb{S},\cos\left(2\phi_{q_T}\right)}}{\mathcal{F}_{UU}^{\mathbb{S}}}\,,\qquad \mathcal{A}_{\rm DIS}^{\rm Sivers}=\frac{\mathcal{F}_{UT}^{\sin\left(\phi_{q_T}-\phi_s\right)}}{\mathcal{F}_{UU}}\,.
\end{align}
For the Collins asymmetry, we use unpolarized TMD PDFs $f_1$, transversity $h_1$ and Collins function $H_1^\perp$ fitted in~\cite{Kang:2015msa}. Although the subset with $\mathbb{S}=\{\pi^+,\ \pi^-\}$ is suppressed due to the sum rule, one can have sizable asymmetries when choosing a subset of either positively or negatively charged pions in {{EEC}}. In the right panel of Fig.~\ref{fig:asyDIS}, we plot the Sivers asymmetry with all the final pions summed over. Here the Sivers functions we adopted were extracted in~\cite{Echevarria:2020hpy}. The asymmetry reaches a few percents at small $\tau$ limit, indicating this observable as a promising tool for further study of the Sivers function.

\section{Conclusion}\label{sec:con}
In this work, we generalize the conventional EEC observable to include the azimuthal angle dependence, where a new EEC jet function referred to as the Collins-type jet function arises. We study the application of this azimuthal angle dependent EEC in both $e^+e^-$ and semi-inclusive deep inelastic scattering, and find that they allow us to explore 3D structure of the nucleon, especially the spin-dependent ones. Thus, the azimuthal angle dependent EEC provides a novel direction for studying the 3D structure of nucleons.

\bibliographystyle{JHEP}
\bibliography{jet.bib}

\end{document}